\documentclass[apj]{emulateapj}
\usepackage{times}
\usepackage{amssymb}
\usepackage{amsmath}
\usepackage{pifont}
\usepackage{graphics}
\usepackage{epsfig}
\usepackage{threeparttable}

\begin{document}

\title{Rotational Period Measurement of Planet Host HD 147379}

\author{Joshua Pepper$^{1}$
}

\affil{$^{1}$Department of Physics, Lehigh University, 16 Memorial Drive East, Bethlehem, PA 18015, USA}

\begin{abstract}

The star HD 147379 was recently announced as a host of a low-mass exoplanet by \citet{Reiners:2017}.  That announcement included a discussion of the possible detection of a periodic signal in the radial velocity observations around 21.1 days, which was interpreted as rotational variability of the star.  We have examined the light curve of this star as observed by the KELT transit survey, and we find a significant periodic signal in the light curve of 22.0064 days, supporting the interpretation of \citet{Reiners:2017} of the rotational variation of the star.

\end{abstract}

\section{Introduction} \label{sec:intro}

In a recent announcement, the CARMENES radial-velocity (RV) search for exoplanets described a discovery of a exoplanet orbiting orbiting HD 147379, a nearby ($\sim$ 11pc) M1Ve star, which shows strong magnetic activity indicators (H$\alpha$ and X-ray emission) \citep{Reiners:2017}.  The exoplanet was detected with an orbital period of $86.54\pm0.06$ days, a minimum mass of $ m\sin{i} = 25 \pm 2 M_{\Earth}$, and located in the conservative habitable zone of its star.  In addition, \citet{Reiners:2017} describe a detection of a periodic signal in their RV data with a period of about 21.1 days, which they interpret as due to the rotational modulation of the host star.  

As a potential explanatory system with a relatively bright host star and a planet in or near the habitable zone, this system will be of great interest to the scientific community.  To help characterize the system, we have examined photometric observations of this star from the KELT exoplanet survey.  We describe our detection of a likely photometric counterpart to the spectroscopic signal indicating the orbital modulation of the star.

\section{Data} \label{sec:data}

The KELT survey is a small-aperture survey of most of the sky, searching from transiting exoplanets orbiting relatively bright ($ 7 < V < 10$) stars \citep{Pepper:2007}.  The operations of the KELT telescopes and the data acquisition and reduction process is described in \citet{Siverd:2012}.  KELT has a 26$\degr$ $\times$ 26$\degr$ field of view.

HD 147379 is at $\alpha = 16^{h} 16^{m} 42.746^{s}$, $\delta = +67\degr 14\arcmin 19\farcs$83 J2000.  It is located in KELT-North field 22, which is centered on $\alpha =$ 16$^{h}$ 03$^{m}$ 00$^{s}$, $\delta =$ +57$\degr$ 00$\arcmin$ 00$\arcsec$ J2000 and was observed 3304 times from UT 2012 Feb 21 to UT 2014 Dec 31.  We combined the TFA-detrended KELT light curves in both east and west orientation (see \citet{Siverd:2012} for details about the KELT data).

\section{Analysis and Results} \label{sec:analysis}

\begin{figure}[ht]
\centering\epsfig{file=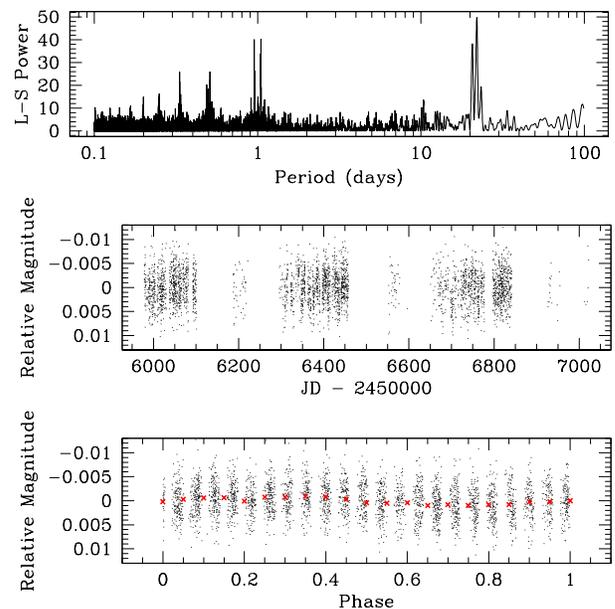,clip=,width=0.99\linewidth}
\caption{{\bf Top:} Lomb-Scargle periodogram of the KELT light curve of HD 147379.  {\bf Middle:} Unphased KELT light curve of HD 147379.  {\bf Bottom:} Light curve phased to the peak of the periodogram, at 22.0064 days.  Black points are individual KELT observations, and the red crosses are binned data at 0.05 in phase.}
\label{fig:fig}
\end{figure}

We searched for periodicity in the KELT light curve of HD 147379 with a generalized Lomb-Scargle \citep[L-S --][]{Press:1992,Zechmeister:2009} analysis, using the VARTOOLS software package \citep{Vart:2016}.  We search a range of periods from 0.1 to 100 days, and find a strong peak at 22.0064 days.  Figure \ref{fig:fig} displays the Lomb-Scargle periodogram, along with the unphased light curve and the light curve phased at the identified period. 

The KELT photometry therefore supports the interpretation from \citet{Reiners:2017} that HD 147379 has a rotational period of 22 days.  The KELT survey has observed a large fraction of the sky for many years, and detected a large number of variable stars.  We provide a list of variable objects in \citet{Oelkers:2017}.  That catalog includes a large fraction of the existing KELT data, although is does not yet include HD 147379.  We expect that an update to that catalog in spring 2018 to include that data.

Furthermore, HD 147379 is at an ecliptic latitude of +79.9 degrees. It therefore falls in the northern continuous viewing zone of upcoming Transiting Exoplanet Survey Satellite (TESS) mission, whose 351-day observing window will cover several phases of this planet's orbital period as well as over 15 phases of its host's rotational period.

\bibliographystyle{apj}
\bibliography{main}

\end{document}